    \documentclass[fleqn,usenatbib]{mnras}

    \usepackage{newtxtext,newtxmath}

    \usepackage[T1]{fontenc}

    \DeclareRobustCommand{\VAN}[3]{#2}
    \let\VANthebibliography\thebibliography
    \def\thebibliography{\DeclareRobustCommand{\VAN}[3]{##3}\VANthebibliography}

    \usepackage{graphicx}	
    \usepackage{amsmath}	
    \usepackage{float}
	\usepackage{hyperref}
	\usepackage[flushleft]{threeparttable}
    \usepackage{booktabs}

    \title[4U $1957\!+\!115$]{Spectro-polarimetry of 4U 1957+115 with IXPE: Effects of spin and returning radiation on polarised emission of black hole in thermal state}

    \author[Ankur Kushwaha et al.]{
	Ankur Kushwaha$^{1,\,2}$\thanks{E-mail: ankurksh@ursc.gov.in},
	Kiran M. Jayasurya$^{1}$,
    Anuj Nandi$^{1}$
	\\
	$^{1}$Space Astronomy Group, ISITE Campus, U. R. Rao Satellite Centre,
	Outer Ring Road, Marathahalli, Bangalore, 560037, India\\
	$^{2}$Department of Physics, Indian Institute of Science, Bangalore, 560012, India\\
	}

    \date{Accepted XXX. Received YYY; in original form ZZZ}

    \pubyear{2023}

\begin{document}
    \label{firstpage}
    \pagerange{\pageref{firstpage}--\pageref{lastpage}}
    \maketitle

	\begin{abstract}
	We present a comprehensive spectro-polarimetric study of persistent 
    Black hole X-ray binary 4U $1957\!+\!115$ with {\it IXPE} and 
    {\it NICER} observations. The source is observed in 
    disk dominated thermal state with disk temperature, 
    kT$_{in}\approx1.4$ keV. 
    The emission during thermal state from the source is 
    found to be moderately polarized and \textit{IXPE} 
    measures a degree of polarization 
    (PD) $= 1.95\pm0.37\%$ ($>4.79\sigma$) along with a polarization angle
    (PA) $= -42.78^{\circ}\pm5.41^{\circ}$
    in the energy range of $2-8$ keV. PD is found to be 
    an increasing function of energy, whereas PA 
    indicates switching within the energy range
    which could be due to high inclination and the 
    returning radiation within the system.
    Simultaneous energy spectra ($0.6-10$ keV) from {\it NICER} 
    are modelled to 
    study the spectral properties. Furthermore, the spin parameter 
    of the black hole is estimated with spectro-polarimetric data 
    as a$_{\ast}=0.988\pm0.001\,(1\sigma)$, which is corroborated by {\it NICER} 
    observations. Finally, we discuss the implications of our findings. 
	\end{abstract}


    \section{Introduction}
	\label{sec: intro}

    The X-ray polarimeter experiments are now detecting
    and investigating the polarimetric properties of bright
    galactic and extra-galactic X-ray objects. The Black hole 
    X-ray binaries (BH-XRBs) are among the suitable candidates 
    for X-ray polarimetric studies as they are bright 
    and are predicted to emit polarised radiation
    \citep{1975MNRAS.171..457R, 1975ApJ...198L..73L,
    1976ApJ...203..701L, 1977Natur.266..429S, 1977Natur.269..128C}.
    
    The polarimetric measurements are combined with simultaneous 
    spectroscopic data for holistic approach to understand
    the emission processes in the vicinity of black hole. 
    The spectro-polarimetric studies are also effective 
    in probing the environment and accretion disk structure, which is 
    formed surrounding the black hole with the matter accreted from the
    companion star. The Keplerian accretion disk is the main source of 
    multi-coloured thermal radiation \citep{1973A&A....24..337S,
    1973blho.conf..343N}, while a Comptonizing medium is believed as the origin
    of the non-thermal component \citep{1994ApJ...434..570T, 1995ApJ...455..623C},
    typically present, in the spectra of different spectral states of BH-XRBs 
    \citep[and references therein]{2005Ap&SS.300..107H, 
    2006ARA&A..44...49R, 2012A&A...542A..56N, 
    2020MNRAS.497.1197B, 2021MNRAS.507.2602K}.

    In recent years, the Imaging X-ray Polarimetry Explorer
    (IXPE; \citet{2022JATIS...8b6002W} has provided high quality
    long exposure observations of a few persistent and transient
    BH-XRBs during different spectral states. The spectro-polarimetric 
    studies have been reported for Cyg X-1 \citep{2022Sci...378..650K}, 
    Cyg X-3 \citep{2023arXiv230301174V}, 4U $1630\!-\!47$ 
    \citep{2023MNRAS.524L..15K, 2023ApJ...949L..43R, 
    2023arXiv230412752R, 2023arXiv230510630R}, LMC X-1 
    \citep{2023arXiv230312034P} and
    LMC X-3 \citep{2023arXiv230906845M, 2023arXiv230910813S}.
    Polarimetric properties of Cyg X-1 in low/hard state (LHS)
    and of 4U $1630\!-\!47$ in high/soft state (HSS) as well as 
    in steep power law state (SPL) are reported to be inconsistent
    with the existing models \citep{2008MNRAS.391...32D, 2009ApJ...701.1175S, 2020MNRAS.493.4960T}.
    These sources exhibit higher degree of polarization (PD)
    with unexpected energy dependence of polarization angle (PA),
    while the polarimetric observables of LMC X-1 and LMC X-3
    in HSS are inline with the ones anticipated from the systems.
        
    Moreover, the X-ray emission, during HSS, of BH-XRBs 
    deeply affected by the spin of the BH and returning 
    radiation that causes depolarization which results
    into reduction of net PD and rotation of PA
    \citep{1980ApJ...235..224C, 2009ApJ...701.1175S}. Hence, 
    a BH-XRB with rapidly spinning BH 
    is expected to exhibit lower PD than that of a system 
    with a slow spinning BH (larger inner disk radius). 
    In order to verify these effects, we conduct a spectro-polarimetric 
    study of 4U $1957\!+\!115$ combining polarimetric data of 
    \textit{IXPE} observations with simultaneous spectral 
    data of \textit{NICER}.

    4U $1957\!+\!115$ is a persistent low mass X-ray binary system 
    harbouring a black hole \citep{1978ApJ...221..907M, 
    1987ApJ...312..739T, 1999MNRAS.306..701H}.    
    It is located in galactic halo (away from galactic plane) 
    and has a low absorption column density of 
    N$_{\rm H}\sim\!0.1\times10^{22}$ cm$^{-2}$ 
    \citep{1993MNRAS.264..411Y,1999ApJ...522..476N}.
    The estimation of mass (M$_{\rm{BH}}$), 
    distance (D) and inclination ($i$) of the system
    have been attempted in multiple studies based on 
    optical and X-ray, but are not very definitive. 
    The best estimation 
    of these parameters are M$_{\rm{BH}}<\!10$ M$_{\odot}$, 
    D $\leq30$ kpc and high inclination of 
    i$\sim\!70^{\circ}-75^{\circ}$ 
    \citep{1999MNRAS.306..701H, 2011ApJ...730...43B,
    2012AJ....144..108M, 2015ApJ...809....9G, 
    2020MNRAS.498L..40M, 2021RAA....21..214S, 2023ApJ...944..165B}. 
    X-ray observations also suggest that 
    the source hosts a rapidly rotating BH
    \citep{1999ApJ...522..476N, 2008ApJ...689.1199N,
    2012ApJ...744..107N, 2021RAA....21..214S, 2023ApJ...946...19D}
    In a recent study, with \textit{NICER} and \textit{NuSTAR} 
    the spectral fits result spin parameter: $a_{\ast}$ = $0.9961\pm{0.0003}$ 
    \citep{2023ApJ...944..165B}. 
    
    4U $1957\!+\!115$, till date, has remained in disk dominated HSS 
    since its detection in $1973$ \citep{1993MNRAS.264..411Y, 
    1999ApJ...522..476N, 2014ApJ...794...85M, 2012ApJ...744..107N, 
    2020MNRAS.498L..40M, 2022MNRAS.517.4489M} and emits persistent 
    flux in X-rays which is comparable to that of LMC X-1 and LMC X-3
    \citep{2022AdSpR..69..483B}. Moreover, detailed polarization 
    measurement during HSS of 4U $1957\!+\!115$, to the best of 
    our knowledge, has not been reported so far. 
        
    In this work, we focus on the spectro-polarimetric study 
    of 4U $1957\!+\!115$.  
    We make use of {\it IXPE} ($2-8$ keV) and {\it NICER} ($0.6-10$ keV) 
    data to study simultaneous polarimetric and spectral
    properties of 4U $1957\!+\!115$. The spin of BH is also 
    estimated with spectral and spectro-polarimetric modelling.
 
	In $\S$\ref{sec: obs_data}, we provide details of the observations 
	and the steps for data reduction. The polarimetric properties 
    and spectro-polarimetric data modelling using both {\it NICER} as well as 
    {\it IXPE} are presented in $\S$\ref{sec: modelling and results}.
	Finally, we discuss the results and conclude in 
	$\S$\ref{sec: discussion}.    

	\section{Observations and Data Reduction}
 	\label{sec: obs_data}

    \begin{table}
    \centering
    \caption{Log of {\it NICER} and {\it IXPE} observations of 4U $1957\!+\!115$
    considered in the analysis.
			From left to right,
            (1) observation identifier;  
			(2) observation ID;
            (3) MJD of observation;
            (4) exposure time.}
	\label{tab: obs_table}
    \scalebox{1}{
    \begin{tabular}{cccc}
    \hline
    \hline
    Obs Name & ObsID& Start MJD & Exposure (ksec)   \\ 
    \hline
    \multicolumn{4}{c}{NICER}\\
    \hline
    N1  & $6100400102$ & $60078$ & $\sim\!2.9$ \\ 
    N2  & $6100400103$ & $60079$ & $\sim\!4.6$ \\ 
    N3  & $6100400104$ & $60080$ & $\sim\!8.0$ \\ 
    N4  & $6100400105$ & $60081$ & $\sim\!5.1$ \\ 
    N5  & $6100400106$ & $60082$ & $\sim\!11.7$ \\ 
    N6  & $6100400107$ & $60083$ & $\sim\!3.9$ \\ 
    N7  & $6100400108$ & $60084$ & $\sim\!5.0$ \\ 
    N8  & $6100400109$ & $60085$ & $\sim\!8.2$ \\ 
    N9  & $6100400110$ & $60086$ & $\sim\!2.7$ \\ 
    N10 & $6100400111$ & $60087$ & $\sim\!4.2$ \\ 
    N11 & $6100400112$ & $60088$ & $\sim\!2.8$ \\ 
    \hline
    \multicolumn{4}{c}{IXPE}\\
    \hline
     X1 & $02006601$  & $59814$ & $\sim\!572$ ($\sim15$ days) \\ 
    \hline
    \end{tabular}}
    \end{table}

    \begin{figure}
    \centering
	    \includegraphics[trim={0 2mm 0 0}, clip,width=\columnwidth]{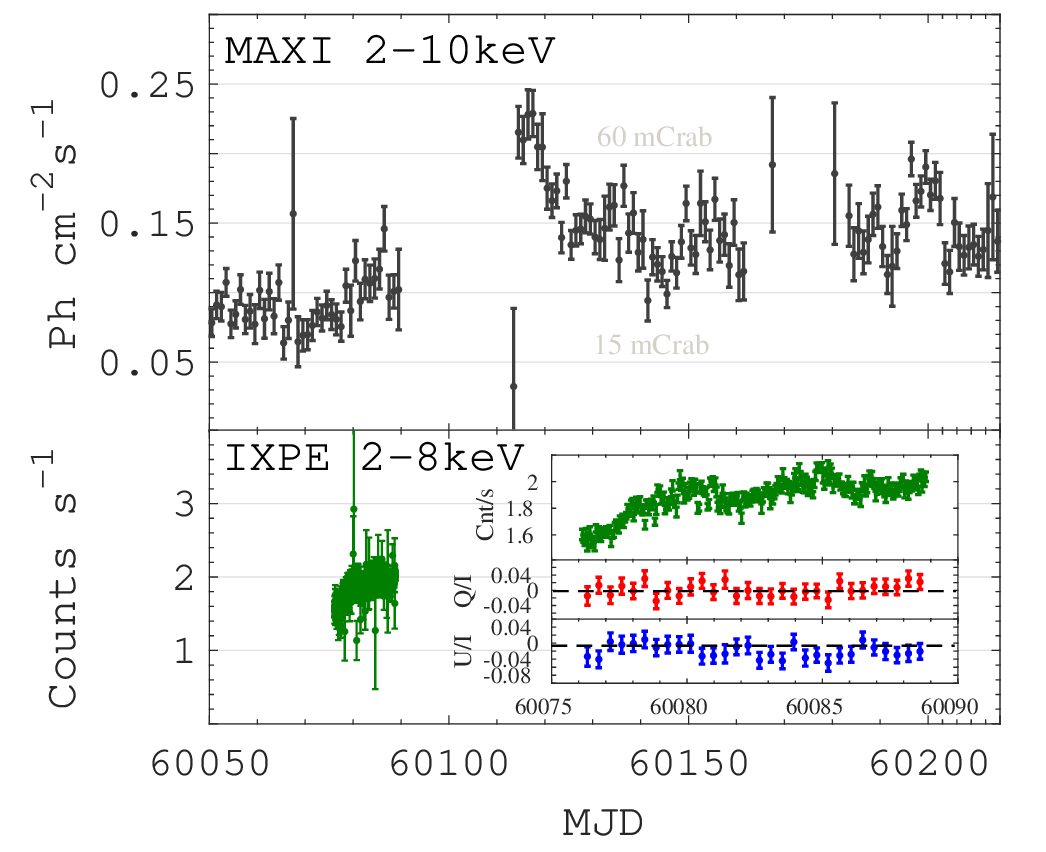}
        \caption{Top to bottom: The light curve of 4U $1957\!+\!115$  
        obtained from {\it MAXI} (purple; 1 day averaged )
        and all three DUs of {\it IXPE} (green; 2 ksec bin time). 
        The inset in the bottom panel shows {\it IXPE} light curve (4 ksec bin time) 
        along with the temporal variation of normalized Stokes parameters Q/I and U/I
        ($\sim$36 ksec bin time). The corresponding energy ranges of light curves are also 
        mentioned in each panel. See the text for details.}
	\label{fig: all_lightcurve}	
	\end{figure}

    \subsection{IXPE}
    {\it IXPE} is an imaging polarimeter \citep{2022JATIS...8b6002W} 
    consisting of three detector units (DUs) which are 
    polarization sensitive in the $2-8$~keV energy range. 
    It observed 4U $1957\!+\!115$ from $12$ May, $2023$ 
    (MJD $60076$) to $26$ May, $2023$ (MJD $60090$) for
    $\sim\!572$ ksec (\autoref{tab: obs_table}).
    The Level-2 data of the observation are reduced and analysed 
    with {\tt IXPEOBSSIMv30.0.0} \citep{2022SoftX..1901194B}
    following the procedure of \citet{2023MNRAS.519.3681F}. 
    Further, {\tt XPSELECT} task is used to  extract the source 
    and background event lists. The source is considered within a 
    circular region of $80"$ and the background region as an annular region with 
    inner \& outer radii of $180"$ \& $240"$, respectively. 
    Subsequently, the {\tt XPBIN} task is
    used to generate the polarization cubes using the {\tt PCUBE} algorithm. 
    The Stokes I, Q and U spectra of source and background are generated
    with the {\tt PHA1}, {\tt PHA1Q} and {\tt PHA1U} algorithms.
    The light curves in $2-8$ keV energy band for the different 
    DUs are generated with the {\tt XSELECT} task of {\tt HEASOFT v6.31.1} 
    and added using the {\tt lcmath} task to get a 
    combined light curve (see bottom panel of \autoref{fig: all_lightcurve}).

    \subsection{NICER}
    {\it NICER} observed the source from $12$ May, $2023$ (MJD $60076$) 
    to $24$ May, $2023$ (MJD $60088$) which overlaps with 
    {\it IXPE} observations (see \autoref{tab: obs_table} and 
    \autoref{fig: all_lightcurve}). 
    We make use of eleven observations of during this interval 
    are analyzed for this work.
    The {\tt NICERDASv10} software distributed 
    with {\tt HEASOFT v6.31.1} is used along with the latest 
    {\tt CALDB} to reduce the data from the observations. The {\tt nicerl2} 
    task is used to perform standard calibration and screening to 
    generate cleaned event lists. The source and background spectra 
    along with the responses are generated in the $0.5-12$ keV energy 
    band using the {\tt nicerl3-spect} task. The spectra 
    are rebinned to have a minimum of $25$ counts per energy bin for
    spectral modelling. 
    

    \section{Modelling and Results}
    \label{sec: modelling and results}
    \subsection{Polarimetric Properties}
    {\it IXPE} observes 4U $1957\!+\!115$ for  $\sim\!572$ ksec
    which spanned over $\sim\!15$ days (see \autoref{tab: obs_table}).  
    The \textit{IXPE} light curve is shown in \autoref{fig: all_lightcurve}.
     The source emanates a persistent flux during the exposure
     with slight increase in count rate. The averaged count rate (from 3 DUs)
    increased from about 1.6 cts/s to 2 cts/s at the start and end of 
    the observation, respectively. The temporal variation of the 
    polarization observables: normalized Stokes  parameters (Q/I \& U/I) 
    are also shown in the \autoref{fig: all_lightcurve} (Inset of bottom panel). We note that the variability is within statistical limits.
    \begin{figure}
		\includegraphics[trim={0 8mm 0 10mm}, clip,width=\columnwidth]{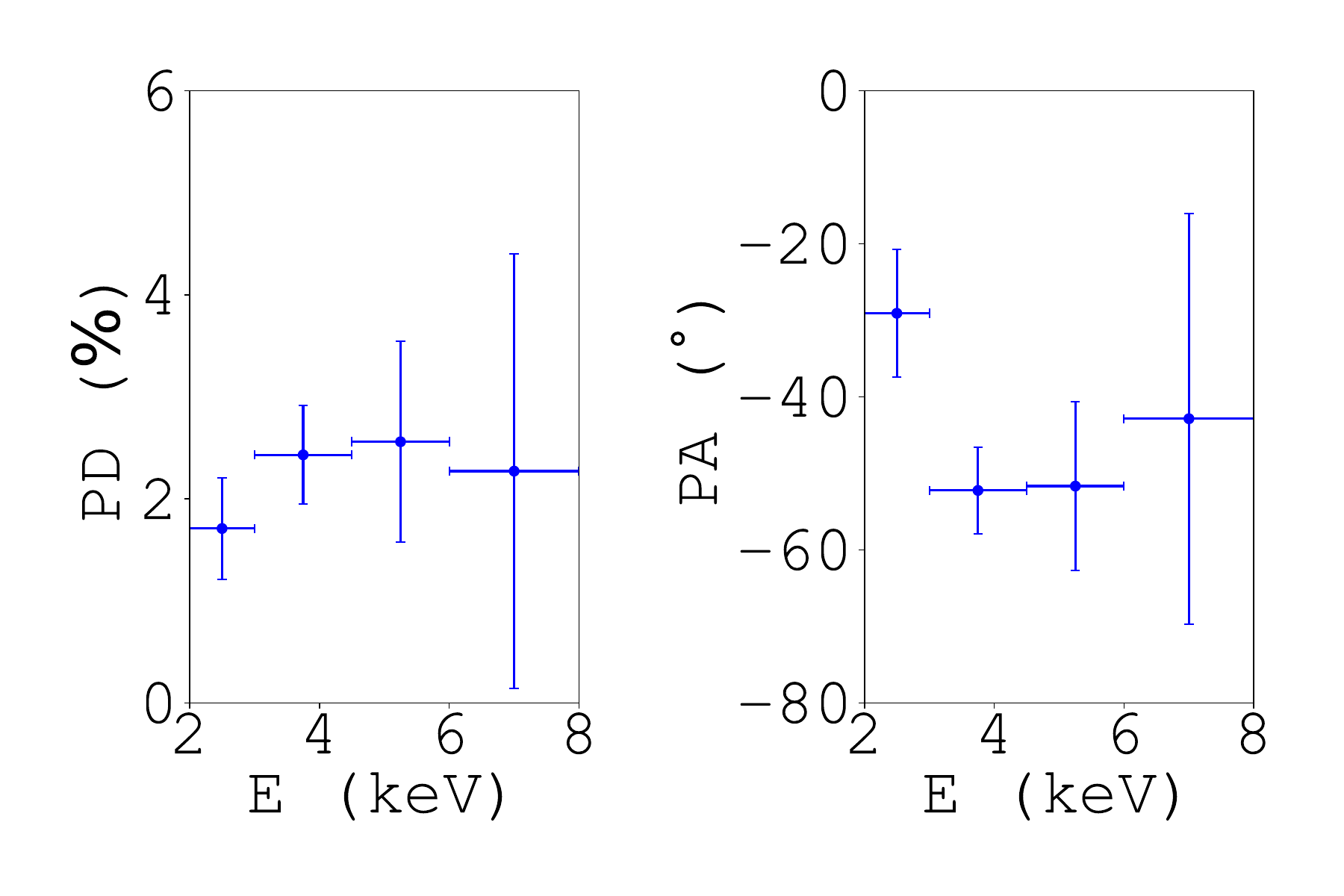}
	    \caption{Polarization degree and associated polarization angle measured from 4U $1957\!+\!115$
        with \textit{IXPE} observations. See the text for details.}
	    \label{fig: ixpe_pa_evar}
	\end{figure}
    In order to investigate polarimetric properties for the first time 
    from this source, we adopt the methodology of \citet{2023MNRAS.524L..15K, 2023MNRAS.521L..74C,
    2023MNRAS.525.4657J}. Firstly, 
    the model-independent {\tt PCUBE} algorithm 
    \citep{2015APh....68...45K} is used to determine  
    the normalized Stokes parameters (Q/I \& U/I) and then polarization 
    degree (PD) and polarization angle (PA) are derived in the 
    $2-3$ keV, $3-4.5$ keV, $4.5-6$ keV, $6-8$ keV 
    and in the total $2-8$ keV energy ranges.
    The resulting parameters along with corresponding MDP 
    and statistical significance are provided in \autoref{tab:pcubetable}.
    The polarization measurements are above the MDP and significant 
    upto 4.5 keV energy range. The measurements in higher energy range
    are less reliable. Alternatively, we also derived PD in
    $2-4$ and $4-8$ keV energy ranges as  $1.58\pm0.36\%\,(>3.75\sigma$; 1.10\% MDP$_{99}$) 
    and $3.26\pm0.76\%\,(>3.67\sigma$; 2.33\% MDP$_{99}$), respectively.
    \begin{figure}
    \hskip -3 mm
		\includegraphics[trim={10mm 25mm 10mm 25mm}, clip,width=1.1\columnwidth]{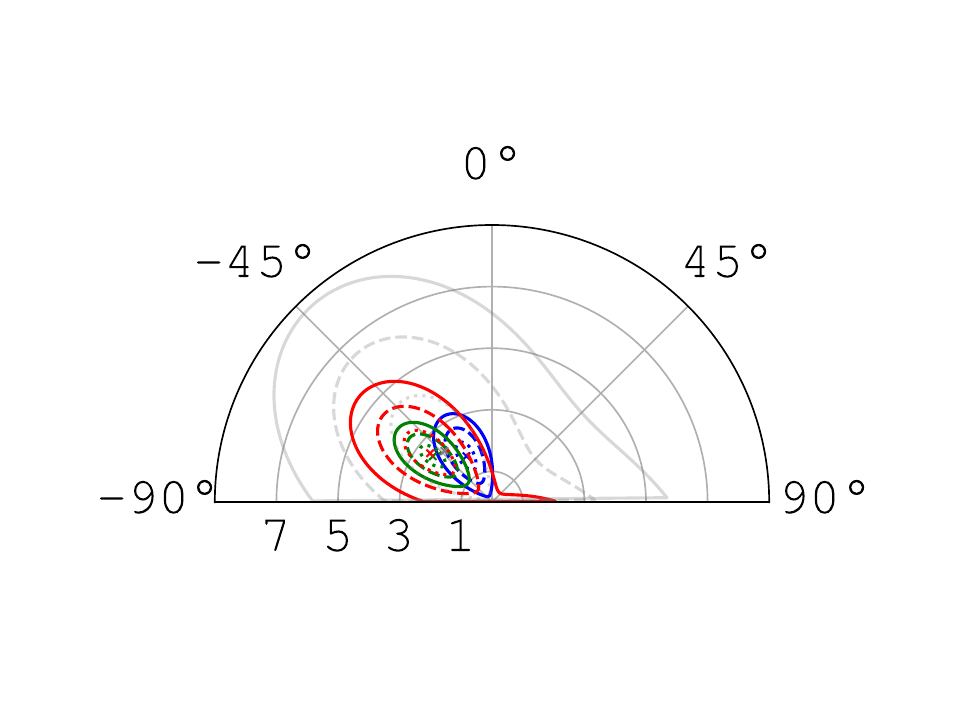}
	    \caption{The confidence contours ($1\sigma$ [dotted], $2\sigma$ [dashed] and 
        $3\sigma$ [solid]) of PA and PD obtained with \texttt{PCUBE} 
        in $2-3$ (blue), $3-4.5$ (green), $4.5-6$ (red) and $6-8$ keV (grey) energy bins. 
        The grid shows PA ($^{\circ}$) on radial lines and PD (\%) as concentric rings. 
        See the text for details.}
	    \label{fig: contour_plot}
	\end{figure}
     These measurements evidently suggest an increasing nature of PD with energy.
    PA is noted to be rapidly declining from $\sim-30^\circ$ to $\sim-50^\circ$.
    The variation of PD and PA with energy is shown in 
    \autoref{fig: ixpe_pa_evar}.
    \begin{table*}
    \centering
    \caption{Polarization parameters obtained using the \texttt{PCUBE} algorithm (for all 3 DUs combined) in different energy   bands. The uncertainties are reported at 1$\sigma$ level. The symbol $^{\dagger}$ denotes
    value below zero.}
    \label{tab:pcubetable}
    \begin{tabular}{lccccc}
        \hline
        \hline
        Parameter & $2-3$ keV &      $3-4.5$ keV &       $4.5-6$ keV &          $6-8$ keV & $2-8$ keV\\
        \hline
        Q/I (\%) & 0.9 $\pm$ 0.5 &  -0.61 $\pm$ 0.48 &   -0.59 $\pm$ 0.98 &    0.17 $\pm$ 2.13  & 0.15 $\pm$  0.37\\
        U/I (\%) & -1.45 $\pm$ 0.5 &  -2.36 $\pm$ 0.48 &   -2.49 $\pm$ 0.98 &   -2.27 $\pm$ 2.13 & -1.94 $\pm$  0.37 \\
        PD (\%) &  1.71 $\pm$ 0.5 &   2.43 $\pm$ 0.48 &    2.56 $\pm$ 0.98 &    2.27 $\pm$ 2.13  & 1.95 $\pm$  0.37\\
        PA ($^\circ$) &-29.08 $\pm$ 8.34 & -52.22 $\pm$ 5.67 & -51.66 $\pm$ 10.99 & -42.85 $\pm$ 26.82 & -42.78 $\pm$ 5.41  \\
        \hline
        MDP$_{99}$(\%) &          1.51 &              1.46 &               2.98 &               6.45 & 1.12\\
        Signif.($\sigma$) & 2.78 &              4.54 &               1.83 &              -$^{\dagger}$ & 4.79\\
        \hline       
    \end{tabular}
    \end{table*}

    The measured PD and PA in different energy ranges spanning $2-8$ keV 
    are also plotted with error contours (see \autoref{fig: contour_plot}). 
    We note a PD = $1.95\%\pm0.37\%$ in $2-8$ keV energy band 
    with all the event data from three DUs combined. The PA in the 
    same energy band is found to be $-42.78^{\circ}\pm5.41^{\circ}$.
  
    \subsection{Spectral Modelling}
    \label{sec: diskbb fit}
	We consider $0.6-10.0$ keV range of {\it NICER} data to model 
    the spectra of all the observations (see \autoref{tab: obs_table}) 
    of $4U\,1957+115$. Firstly, we fit a phenomenological model {\tt Tbabs*diskbb} 
    on the set of spectra. All the spectra are fitted at once
    without tying the parameters. The model results in good fits and 
    residual does not suggest any requirement of a power-law 
    towards the high energy in the spectra. It indicates that the 
    source is in a steady disk dominated state during multiple
    observations of \textit{NICER}. The fits give 
    an average disk temperature of kT$_{in}\approx1.4\pm0.01$ keV. 
    The combined fit statistics for $11$ spectra results into a 
    $\chi_{red}^{2}\,(\chi^{2}$/dof)= $0.93$. The hydrogen column 
    density N$_{\rm H}$ 
    is estimated to be $\approx0.85\times$10$^{22}$ atoms cm$^{-2}$ 
    from the group spectral fits. The estimation of N$_{\rm H}$ is 
    consistent with previous estimates from spectral studies
    \citep{1993MNRAS.264..411Y, 
    1999ApJ...522..476N, 2014ApJ...794...85M, 2012ApJ...744..107N, 
    2020MNRAS.498L..40M, 2022MNRAS.517.4489M, 2023ApJ...944..165B}.
    The overall fit of obs. N5 is shown in 
    \autoref{fig: nicer_ixpe_spectra}.
    
    \subsection{Spin Estimation with Continuum Fitting (CF) Method }
    \label{sec: kerrbb fit}
    In $\S$ \ref{sec: diskbb fit}, it is shown that the source 
    exhibits a steady disk dominated soft state during {\it NICER} 
    observations. Further, the CF method is employed and only
    large exposure spectra (Obs. N3, N5 and N8) are modelled 
    with relativistic model to estimate the spin parameter 
    of BH. For relativistic modelling of spectra, 
    we follow \citet{2021MNRAS.507.2602K}
    and replace {\tt diskbb} with {\tt kerrbb} in our previous 
    model mentioned in $\S$\ref{sec: diskbb fit}. 
    Hence, we fit all spectra with {\tt Tbabs*kerrbb}. 
    The model results into reasonably acceptable fits. 
    We, once again, introduce a power-law component 
    {\tt simpl} to check if there is any signature of 
    non-thermal component of corona present in the spectra. 
    We do not observe any improvement in overall fit and notice 
    unacceptable values of parameters in {\tt simpl} component. 
    Therefore, we continue to fit 
    the spectra with our initial {\tt kerrbb} model, which is a 
    relativistic thin accretion disk model and includes self-irradiation 
    along with limb-darkening effects. We switch off limb-darkening and 
    apply zero torque condition at the inner boundary of the disk. 
    Moreover, model is provided with best estimations of distance to 
    source (D), mass of the black hole (M$_{\rm BH}$) in the binary system and 
    inclination of the binary plane ($i$) based on the published works (see $\S$\ref{sec: intro} and references therein). 
    Hence, we fix the distance at D = $16$ kpc, the inclination 
    at $i$ = $75^{\circ}$, black hole mass at M = 8M$_{\odot}$. 
    The spectral hardening factor ($f$) is fixed to a fiducial 
    value of 1.6 \citep{1995ApJ...445..780S}. The other two parameters 
    of {\tt kerrbb} namely accretion rate ($\dot{\rm{M}}$) 
    and spin parameter (a$_{\ast}$) are allowed to vary freely.
    For obs. N5, overall fit estimates N$_{H}$ = $0.11\pm0.01\times10^{22}$ 
    atoms cm$^{-2}$, a$_{\ast}=0.994\pm{0.002}$ and 
    $\dot{\rm{M}}$ = $0.13\pm0.01\,\times10^{18}$ g s$^{-1}$
    with $\chi^{2}_{red}\,(\chi^{2}$/dof) = $0.93(792.5.65/850)$. We highlight the 
    fact that estimation of spin parameter is highly 
    dependent on D, M and $i$. Further, similar fit procedure is 
    applied to rest of the observations to estimate the spin. 
    We notice a consistent value of spin parameter
    from all the selected observation from HSS of 
    4U $1957\!+\!115$. The best fitting model unfolded against the
    spectral data for obs. N5 is shown in \autoref{fig: nicer_ixpe_spectra}. 

   \subsection{Spectro-polarimetric Analysis}
    \begin{figure}
		\includegraphics[trim={0.4cm 0 0 0}, clip,width=1.05\columnwidth]{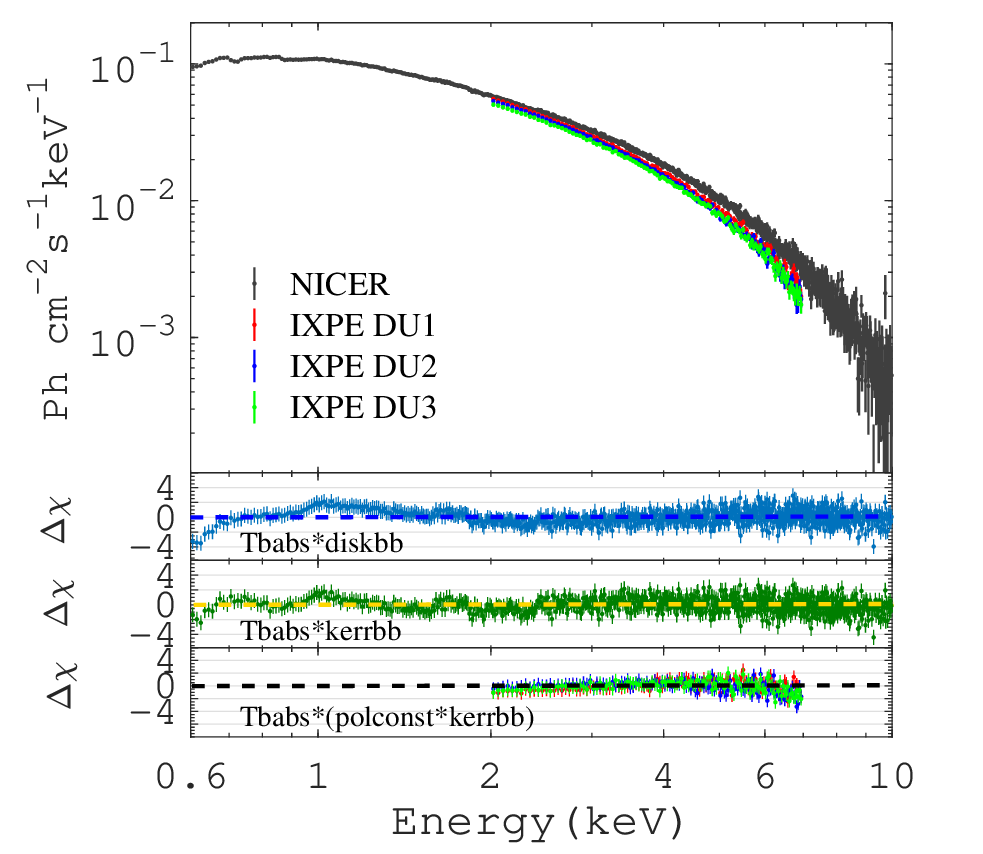}
	    \caption{4U $1957\!+\!115$ spectra (unfolded) obtained from {\it NICER} observations obs. N5 is
     shown along with \textit{IXPE} (DU1, 2 and 3) in top panel. Bottom panels show the residuals 
     (in units of $\sigma$) corresponding to each model considered in this work. See the text for details.}
	    \label{fig: nicer_ixpe_spectra}
	\end{figure}

    Subsequently, we also perform model-dependent analysis by 
    fitting {\it IXPE} Stokes I, Q and U spectra in {\tt XSPEC}. 
    We make use of the relativistic model as described in 
    $\S$\ref{sec: kerrbb fit} and introduce the multiplicative component 
    {\tt polpow} which has PA and PD as function of energy. 
    The component defines PD(E) = A$_{\rm norm}\times$ E$^{-{\rm A}_{\rm index}}$ 
    and PA(E) = psi$_{\rm norm}\times$E$^{-{\rm psi}_{\rm index}}$. The
    {\tt polpow} is chosen for polarimetric component based on
    the results of {\tt PCUBE} algorithm that shows that PD and PA
    are energy dependent. 
    Hence, the model {\tt Tbabs(polpow*kerrbb)}
    is applied to fit three Stokes spectra from all the DUs 
    of {\it IXPE} in {\tt XSPEC}. 
    Only parameters of {\tt polpow} component and 
    a$_{\ast}$, $\dot{\rm{M}}$ of {\tt kerrbb} are 
    kept free for the fit. All other 
    parameters are frozen to average values obtained 
    from {\it NICER} spectral fits (see $\S$\ref{sec: diskbb fit}).
    The fit procedure results into 
    an acceptable $\chi_{red}^{2}$($\chi^{2}$/dof) =$0.93(1103.6/1108)$ for 
    combined nine spectra (I, Q \& U for 3 DUs). 
    Moreover, we replace {\tt polpow} with {\tt polconst}
    to verify the polarization measurements in overall $2-8$ keV
    energy range as well with assuming no energy dependence. 
    The best fit values are provided in the \autoref{tab: fit_table}.
   
    \begin{table}
    \centering
    \caption{The best fit spectral parameters from {\it IXPE} 
    observations of 4U $1957\!+\!115$ with spectro-polarimetric model.
     From left to right are,
			 (1) model components; 
		      (2) parameters values of components with \texttt{polconst};  
			     (3) with \texttt{polpow}. The parameters
            that are fixed  during the fits are denoted with $^{\rm fixed}$.}
	       \label{tab: fit_table}
        \scalebox{0.9}{
    
        \begin{tabular}{ccc} 

		    \hline
			\hline
			Components 		&  \multicolumn{2}{c}{Model}\\
            \cmidrule(lr){2-3}
            & {\tt Tbabs*polconst*kerrbb} & {\tt Tbabs*polpow*kerrbb}\\
			\hline
			  A$_{\rm norm}(\times10^{-2}$)		    &$1.8\pm0.2$ & $0.54\pm0.3$\\
            A$_{\rm index}$		    &- &$-1.08\pm0.3$\\
            psi$_{\rm norm}$		    &$-43.1\pm5.2$&$-14.4\pm6.3$\\
            psi$_{\rm index}$		    &-&$-0.87\pm0.01$\\
            N$_{\rm H}(\times10^{22}$ cm$^{-1}$)  &$0.11^{\rm fixed}$& $0.11^{\rm fixed}$\\
			a$_{*}$					&$0.988\pm0.001$&$0.983\pm0.003$\\
		      $\dot{\rm{M}}(\times10^{18}$ g s$^{-1}$)			&$0.13\pm0.01$&$0.15\pm0.02$\\
			    								
			\cmidrule(lr){2-3}
			$\chi_{red}^{2}$ ($\chi^{2}$/dof)   &$0.93\,(1013.54/1109)$ & $0.91\,(1013.6/1108)$\\
			                     
			\hline                     
		\end{tabular}}
    \end{table}
    The overall fit and modelled spectra are shown in 
    \autoref{fig: nicer_ixpe_spectra}. Furthermore, the model-dependent 
    polarization results are compared with those 
    obtained in model-independent method by 
    integrating PD(E) and PA(E) using {\tt polpow} parameters 
    in the $2-8$ keV energy range. PD and PA are estimated
    to be $2.6\%$ and $\sim-52^{\circ}$, respectively.
    We notice that parameters of {\tt polpow} have
    large uncertainties and poor constraints.
    The {\tt polconst} component parameters result 
    $1.8\%\pm0.2\%$ and $-43^{\circ}\pm5.2^{\circ}$
    for PD and PA respectively.
    We notice measurements of PA and PD obtained from {\tt XSPEC} 
    differ marginally from results of {\tt PCUBE} algorithm. 
    We outline that the spin parameters values resulting from 
    {\it IXPE} (see \autoref{tab: fit_table}) polarimetric fits are inline with the values
    obtained by {\it NICER} spectral fits (see $\S$\ref{sec: diskbb fit}).

    \section{Discussion}
    \label{sec: discussion}
    In this letter, we report the measurements of X-ray 
    polarization from a persistent low mass BH-XRB 4U $1957\!+\!115$
    by {\it IXPE} during HSS. The polarization 
    properties are combined with simultaneous spectral 
    modelling of {\it NICER} observations for 
    spectro-polarimetric study of the source.

    Our finding, based on spectro-polarimetric fits
    with constant polarization component model
    gives measurements of PD$=1.8\%\pm0.2\%$
    and PA$=-43^{\circ}\!\pm5.2^{\circ}$ in $2-8$ keV
    energy range. The model-independent approach
    using {\tt PCUBE} algorithm estimates PD$=1.95\%\pm0.37\%$
    and PA$=-42.78^{\circ}\!\pm\!5.41^{\circ}$.
    These measurements are in good agreement. However,
    spectral modelling with inclusion of an energy-dependent 
    component results into marginally higher values of
    polarization with PD$\sim2.6\%$
    and PA$=-52^{\circ}$ in $2-8$ keV
    energy range.

    Moreover, we notice an increasing nature of PD
    with energy when PD is estimated in different energy
    ranges using \texttt{xpbin} tool (see \autoref{fig: ixpe_pa_evar}). 
    Such energy dependence is previously observed in HSS
    of 4U $1630\!-\!47$ \citep{2023MNRAS.524L..15K, 2023ApJ...949L..43R, 
    2023arXiv230412752R} and LMC-X3 \citep{2023arXiv230906845M, 2023arXiv230910813S}. We compare the PD with the same of
    LMC X-3 which is consistent with existing theoretical models. 
    LMC X-3 and 4U $1957\!+\!115$ have roughly similar high inclination angles
    but the former has a BH with lower spin than the later.
    Thermal disk emission from a low spin high inclination
    BH system is expected to show higher PD than the one with 
    high spin high inclination. A slow spinning BH, where the inner disk radius is large, the polarization angle does not rotate much. For a high spin BH, where the inner disk radius is small, the polarization angle of the photons produced at the inner parts of the disk is much rotated, causing depolarization. Therefore, the net polarization is lower in the case of a higher BH spin \citep[and references therein]{2023MNRAS.519.6138M}.
    This can be related to the cases of
    LMC X-3 and 4U $1957\!+\!115$ as the observed PD is about
    $3\%$ and $2\%$, respectively up to 5 keV. For a comparison
    above 5keV more \textit{IXPE} observation are required
    as the polarimetric results suffer poor statistics near
    higher energies.

    PD is also affected by returning radiation \citep{1980ApJ...235..224C, 2009ApJ...701.1175S} and its albedo \citep{2020MNRAS.493.4960T, 2023MNRAS.519.6138M}
    on the accretion disk. The returning radiation have significant 
    effect on polarization around high spin BH where the inner radius
    of disk is small which causes more returning radiation to scatter
    from the near side of the disk when compared to low spin BH. The
    returning radiation shows up as higher polarisation but in higher
    energies and may not be captured within {\tt IXPE} energy range, 
    but interestingly it also causes switching of PA within $2-8$ keV
    from a BH thermal disk with appropriate albedo. Hence, a low spin 
    BH is expected to have a roughly constant PA while a high spin BH
    will have a switching PA to lower values \citep{2020MNRAS.493.4960T, 2023MNRAS.519.6138M}. We observe a rapidly declining
    PA from about $-30^\circ$ to $-50^\circ$ which could be
    an indication of switching PA in 4U $1957\!+\!115$. Such switching is 
    not present in the case of LMC X-3 which exhibits an anticipated
    constant PA of $\sim-42^{\circ}$.

    We also highlight the fact that measurements of PA for 4U $1957\!+\!115$
    can not be related to orientation of system or disk plane as there
    is no radio activity detected from the source. We note that averaged
    PA measured in $2-8$ keV of 4U $1957\!+\!115$ is comparable to that of
    observed in case of LMC X-3.

    Furthermore, spectro-polarimetric fit of {\it IXPE} data with 
    relativistic model also constrains the BH spin 
    parameter: a$_{\ast}=0.988\pm0.001\,(1\sigma)$ and accretion rate:
    $\dot{\rm{M}}=(0.13\pm0.01)\times10^{18}$ g s$^{-1}$.

    We compare the value of a$_{\ast}$ with that obtained
    by applying CF method on {\it NICER} spectra which 
    resulted a$_{\ast}=0.994\pm0.002\,(1\sigma)$.
    The spin parameter estimated with two different approaches
    are in good agreement. The estimations are also in line
    with findings of \citet{2023ApJ...944..165B}
    Here, we highlight the fact that estimation 
    of spin parameter is highly dependent on $i$, D and M. 

    To summarize, we report measurement of polarized emission from
    4U $1957\!+\!115$ during disk dominated high soft state. The observed degree 
    of polarization and associated angle indicate the effects
    of high spin, high inclination and returning radiation within the system.
    The overall polarisation properties are consistent, at low energies, with 
    expected from existing models. For definitive conclusion in higher energies
    longer exposure polarimetric observations are required.

    \section*{Acknowledgements}
    Authors thank GH, SAG; DD, PDMSA and Director, URSC for 
	encouragement and continuous support to carry out this 
    research. {\it IXPE}, {\it NICER} and {\it MAXI} teams
    are also thanked for providing data products and software tools
    for data analysis.
    
    \section*{Data Availability}
    Data used for this work are available at 
	{\it HEASARC} website 
    (\url{https://heasarc.gsfc.nasa.gov/docs/archive.html})
	and {\it MAXI} website
	(\url{http://maxi.riken.jp/top/index.html}).



    \bibliographystyle{mnras}
    \bibliography{references} 
        


    \bsp	
    \label{lastpage}
    \end{document}